\documentclass[twocolumn,prd,showpacs,preprintnumbers,amsmath,amssymb,epsf,superscriptaddress]{revtex4}

\usepackage{graphicx}
\usepackage{dcolumn}
\usepackage{bm}

\begin{document}

\preprint{FERMILAB-PUB-06-065-T,BNL-06/2}

\title{Higgs Triplets and Limits from Precision Measurements}

\author{Mu-Chun Chen}
\affiliation{%
Theoretical Physics Department, Fermilab, Batavia, IL 60510, USA
}%

\author{Sally Dawson}%
\author{ Tadas Krupovnickas}
\affiliation{%
Department of Physics, Brookhaven National Laboratory,
Upton, NY 11973-5000, USA
}%

\date{\today}

\begin{abstract} 
In this letter, we present our results on a global  fit to precision
electroweak data in a Higgs triplet model. 
In models with a triplet Higgs boson,  a 
consistent renormalization scheme differs from that of the Standard 
Model and the global fit shows 
that a light Higgs boson with mass of $100-200$ GeV is preferred. 
Triplet Higgs bosons arise in many extensions of the Standard Model, including 
the left-right model and the Little Higgs models.  
Our result demonstrates the importance of the scalar loops when there is a large mass 
splitting between the heavy scalars. It also indicates the significance of the global fit. 

\end{abstract}

\maketitle

{\bf 1.~Introduction: }
One of the major goals of the LHC is uncovering the mechanisms of 
electroweak symmetry breaking and the generation of fermion masses. In the 
Standard Model (SM) of particle physics, the masses of gauge bosons and fermions 
are generated by the interactions with a single scalar field. 
After  spontaneous symmetry breaking, a neutral CP-even 
Higgs boson, $h$, remains as a physical particle and the fermion 
and gauge boson masses 
arise through couplings to the Higgs boson. Discovering the Higgs particle and 
measuring its properties is central to an understanding of 
electroweak symmetry breaking.  

Measurements at LEP, SLD, and the Tevatron have been  extensively used  to 
restrict the parameters of the Standard Model. 
In the SM, the mass of the Higgs boson is strongly constrained by 
precision electroweak measurements. If there are new particles or new 
interactions beyond those of the SM, a global fit to the experimental data 
can yield information about the allowed parameters of the model.

In models which contain more Higgs bosons than the $SU(2)_L$ doublet of the 
SM, there are more parameters in the gauge/Higgs sector than in the Standard 
Model.  If these additional Higgs bosons are in $SU(2)_L$ 
representations other than singlets and doublets, the 
SM relation, $\rho=M_W^2/(M_Z^2 \cos^2\theta_W)=1$ does not 
hold at tree level. This has the implication that when the theory is 
renormalized at one-loop, extra input parameters beyond those of the 
SM are required~\cite{Lynn:1990zk,Blank:1997qa,Czakon:1999ue,Erler:2004nh,Chen:2003fm,Chen:2005jx}. 

In this letter we consider a simple model with $\rho\ne 1$ at tree level, 
the Standard Model with a Higgs doublet and an additional Higgs triplet. 
Higgs triplets are an essential ingredient of the Little Higgs (LH) class of 
models and so have received significant attention recently~\cite{lh}. 

LH models~\cite{lh} 
are a new approach to understanding
the hierarchy between the TeV scale of possible
new physics and the 
electroweak scale, $v=246~GeV=(\sqrt{2} G_F)^{-1/2}$.
   These models have an expanded 
gauge structure at the TeV scale which contains the Standard Model
$SU(2)\times U(1)$ electroweak gauge groups.  
The LH models are constructed
such that an approximate global symmetry prohibits the Higgs boson from 
obtaining a quadratically divergent mass until at least two loop order.
The Higgs boson is a pseudo-Goldstone 
boson 
resulting from the spontaneous breaking of the
approximate global symmetry and so 
 is naturally light.  The Standard Model then emerges
as an effective theory which is valid below the scale $f$
associated with the spontaneous breaking of the global symmetry. 
LH models contain weakly coupled TeV scale gauge bosons 
from the expanded gauge structure, which
couple to the Standard Model fermions.  In addition, these
new gauge bosons typically mix with the Standard Model $W$ and
$Z$ gauge bosons.  Modifications of the electroweak
sector of the theory, however, are severely restricted by precision
electroweak data and  require the scale of the little Higgs
physics, $f$, to be in the range 
$f>1-6~TeV$~\cite{Csaki:2002qg}, 
depending on the specifics of the model. The LH models also contain expanded Higgs sectors with
additional Higgs doublets and triplets, as well as a new charge $2/3$
quark, which have important implications for precision electroweak
measurements~\cite{Chen:2003fm}. In Ref.~\cite{Chen:2003fm} we found that 
by including the one-loop contributions from the heavy scalars, the scale $f$ can be 
lowered. We have also observed the non-decoupling behaviour of the triplet. 
The non-decoupling of the scalar fields in models with additional scalar fields that acquire 
electroweak breaking VEV was first pointed out in 
Ref.~\cite{Toussaint:1978zm}.

The effects of the prediction for the $W-$boson mass in a Higgs triplet model were 
considered in Ref.~\cite{Chen:2005jx} and here we present a global fit to the electroweak 
observables in this model.

{\bf 2. The Triplet Model: }
We consider the Standard Model with an additional Higgs 
boson which transforms as a real triplet 
under the $SU(2)_L$ gauge symmetry. The $SU(2)_L$ Higgs doublet is identical 
to that of the SM, 
\begin{equation}
H=\left(
\begin{array}{c}
\phi^{+}\\
{1\over \sqrt{2}}(v+\eta^0+i\phi^0) 
\end{array}
\right) \; ,
\end{equation}
while the real triplet is 
\begin{equation}
\Phi=\left(
\begin{array}{cc}
\eta^{+}
\\
v^{\prime}+\eta^{0} 
\\
\eta^{-}
\end{array}\right) \; .
\end{equation}
The $W$ boson mass is given by, 
\begin{equation}
M_W^{2}={g^2\over 4}(v^2+v^{\prime2}) \; ,
\end{equation}
leading to the relationship $v_{SM}^2=(246~\mbox{GeV})^2=v^2+v^{\prime 2}$.

There are four physical Higgs bosons in the spectrum: two 
neutral Higgs bosons, $H^0$ and $K^0$, and a charged 
Higgs boson, $H^\pm$.  The mixing between the two neutral Higgs bosons 
is described by an angle $\gamma$, 
\begin{eqnarray}\label{neutralmix} 
\left(\begin{array}{c} 
H^{0}\\ K^{0} 
\end{array}\right) 
& = & 
\left(\begin{array}{cc} 
c_{\gamma} & s_{\gamma}\\
-s_{\gamma} & c_{\gamma}
\end{array}\right) 
\left(\begin{array}{c}
\phi^{0}\\ \eta^{0}
\end{array}\right)\; .
\end{eqnarray}
 The charged Higgs bosons $H^{\pm}$ are  
linear combinations of the charged components in 
the doublet and the triplet, with a mixing angle $\delta$, 
\begin{eqnarray}
\left(\begin{array}{c}
G^{\pm}\\ H^{\pm}
\end{array}\right)
& = &
\left(\begin{array}{cc}
c_{\delta} & s_{\delta}\\
-s_{\delta} & c_{\delta}
\end{array}\right) 
\left(\begin{array}{c}
\phi^{\pm}\\ \eta^{\pm}
\end{array}\right)\; ,
\end{eqnarray}
where $G^{\pm}$ are the Goldstone bosons corresponding 
to the longitudinal components of $W^{\pm}$. 

In terms of the custodial symmetry violating parameter, $\rho$, the relation 
between the $W$ and $Z$ boson masses is modified from the SM relationship,
$\rho=1$, to be, 
\begin{equation}
\rho={M_W^2\over M_Z^2 \cos^2\theta_W}={1\over \cos^{2} \delta} \, .
\end{equation}
where $v^{\prime} = \frac{1}{2} v \tan\delta~$. 

The symmetry breaking in this model is described by the following scalar potential,
\begin{eqnarray}
V(H,\Phi) &=& \mu_{1}^{2} \bigl|H\bigr|^{2} + \frac{1}{2} \mu_{2}^{2} \Phi^{2} 
+ \lambda_{1} \bigl|H\bigr|^{4} + \frac{1}{4} \lambda_{2} \Phi^{4} 
\nonumber \\ && 
+ \frac{1}{2} \lambda_{3} \bigl|H\bigr|^{2} \Phi^{2} 
+ \lambda_{4} H^{\dagger} \sigma^{\alpha} H \Phi_{\alpha}
\; ,
\end{eqnarray}
where $\sigma^{\alpha}$ denotes the Pauli matrices. 
This model has six parameters in the scalar sector, 
$\bigl(\mu_{1}^{2},\mu_{2}^{2}, 
\lambda_{1},\lambda_{2},\lambda_{3},\lambda_{4}\bigr)$. Equivalently, 
we can choose $\bigl(M_{H^{0}},M_{K^{0}},M_{H^{\pm}}, 
v,\tan\delta,\tan\gamma \bigr)$ as the independent parameters. 
Two of these six parameters, $v$ and $\tan\delta$, contribute to the 
gauge boson masses. The six independent parameters in the 
scalar sector, along with the gauge couplings, $g$ and $g^\prime$ 
completely describe the theory. 
We can equivalently choose the muon decay constant, $G_{\mu}$, 
the Z-boson mass, $M_{Z}$, 
the effective leptonic mixing angle, $s_{\theta}\equiv \sin^{2}\theta_{\mbox{\tiny eff}}$, 
and the fine structure constant 
evaluated at $M_Z$, $\alpha(M_Z)$, 
as our input parameters, along with $M_{H^0}$, $M_{K^0}$, and 
$M_{H^\pm}$ and $\tan\gamma$ and the fermion masses, $m_{f}$. 

From the minimization conditions, we obtain, 
\begin{eqnarray}
4\mu_{2}^{2} t_{\delta} + \lambda_{2}v^{2}t_{\delta}^{3}
+2\lambda_{3}v^{2}t_{\delta}-4\lambda_{4}v = 0  
\label{min1}\\
\mu_{1}^{2} + \lambda_{1}v^{2} + \frac{1}{8} 
\lambda_{3}v^{2}t_{\delta}^{2} 
-\frac{1}{2}\lambda_{4}vt_{\delta} =  0    
\; ,\label{min2}
\end{eqnarray}
where $t_{\delta} \equiv \tan\delta$. 
Consider the case when there is no mixing in the neutral sector, 
\begin{equation}
\frac{\partial^{2} V}
{\partial \phi^{0} \partial \eta^{0}}
= \frac{1}{2} \lambda_{3} v^{2} t_{\delta} - \lambda_{4} v
 = 0 \; ,
\end{equation} 
the condition 
$\tan \delta = (2\lambda_{4}/\lambda_{3}v)$ 
then follows. 
In the absence of the neutral mixing, $\gamma=0$, 
in order to take the charged mixing angle $\delta$ to zero 
while holding $\lambda_{4}$ fixed, one thus has to take 
$\lambda_{3}$ to infinity. In other words, for the 
triplet to decouple requires a dimensionless 
coupling constant $\lambda_{3}$ to become strong, 
leading to the breakdown of the perturbation theory. 
Alternatively, the neutral mixing angle $\gamma$ can approach zero 
by taking $\mu_{2}^{2} \rightarrow \infty$ while keeping 
$\lambda_{3}$ and $\lambda_{4}$ fixed. In this case, 
the minimization condition 
implies that the charged mixing angle  
$\delta$ has to approach zero. This  corresponds to the 
case where the custodial symmetry is restored, as the triplet 
VEV vanishes, $v^{\prime} = 0$. 
However, severe fine-tuning is needed to satisfy the 
minimization condition. 
Another way to get $\delta \rightarrow 0$ is to have $\lambda_{4} 
\rightarrow 0$. 
This corresponds to a case in which the model exhibits tree level 
custodial symmetry. So unless one imposes by hand a symmetry to forbid 
$\lambda_{4}$, four input parameters are always needed in the renormalization. 
As the neutral mixing angle, $\gamma$, does not contributes to the gauge boson 
masses, it is assumed to be zero hereafter and thus the scalar sector consists 
only five parameters. Having a non-zero value for $\gamma$ does 
not change our conclusions. The effects on the heavy scalar masses in the presence 
of a non-zero $\gamma$ can be found in Ref.~\cite{Forshaw:2003kh}.

The effective leptonic mixing angle is defined through the vector and 
axial vector parts of the effective $1$-loop 
$Ze\overline{e}$ coupling, $g_{V}^{e}$ and $g_{A}^{e}$, as, 
\begin{equation}
1-4\sin^{2}\theta_{\mbox{\tiny eff}} = \frac{Re(g_{V}^{e})}{Re(g_{A}^{e})} \; ,
\end{equation} 
while the counter term for $\sin^{2}\theta_{\mbox{\tiny eff}}$ is given by~\footnote{Note that a factor of 
$2s_{\theta}c_{\theta}$ is missing in the first term of Eq.~(3.10) in Ref.~\cite{Blank:1997qa}.}, 
\begin{eqnarray}
\frac{\delta s_{\theta}^{2}}{s_{\theta}^{2}} & = & 
\mbox{Re}\bigg\{ \frac{c_{\theta}}{s_{\theta}}
\biggl[ \frac{g_{V0}^{e~2}-g_{A0}^{e~2}} 
{2s_{\theta}c_{\theta}g_{A0}^{e}} \Sigma^{e}_{A}(m_{e}) 
+\frac{\Sigma^{\gamma Z}(M_{Z})}{M_{Z}^{2}} \nonumber \\
&& 
- \frac{g_{V0}^{e}}{2s_{\theta}c_{\theta}}\biggl(
\frac{\Lambda_{V}^{Zee}(M_{Z})}
{g_{V0}^{e}}-\frac{\Lambda_{A}^{Zee}(M_{Z})}{g_{A0}^{e}}\biggr)\biggr] 
\bigg\} \; .
\end{eqnarray} 
Here $g_{V0}^{e}$ and $g_{A0}^{e}$ are the 
tree level vector and axial vector parts of the $Ze\overline{e}$ 
coupling, $\Sigma_{A}^{e}$ is the axial part of the electron self-energy, 
$\Lambda_{V,A}^{Zee}$ are the un-renormalized $Ze\overline{e}$ vertex 
corrections, and $\Sigma^{\gamma Z}$ is the $\gamma-Z$ two point mixing 
function. 
Experimentally, the measured values for these input 
parameters are~\cite{Eidelman:2004wy}, 
$G_{\mu} =  1.16637(1) \times 10^{-5} \; \mbox{GeV}^{-2}$,  
$M_{Z} =  91.1876(21) \; \mbox{GeV}$, 
$M_W =  80.410(32) \; \mbox{GeV}$, 
$\sin^{2}\theta_{\mbox{\tiny eff}} =  0.2315(3)$ and 
$\alpha(M_Z) = 1/128.91(2)$.

We emphasize that the case considered here is different 
from the model considered 
by Chankowski {\it et al.} in~\cite{Chankowski:2006jk}. 
In the present example, a triplet Higgs 
which has a VEV that breaks the electroweak symmetry is present,  
while in the model of Ref.~\cite{Chankowski:2006jk},  
 additional scalar fields except the 
SM Higgs boson acquire  VEVs that break the electroweak symmetry. In their 
case the effects of the additional Higgs multiplets can be decoupled from the 
SM predictions. 

{\bf 3. Global fit: }
Here we present a global fit to the suite of electroweak 
precision measurements shown in Table 1. 
\begin{table}
\begin{center}
\begin{tabular}{ccc}
Observable & \hspace{0.8cm} & Experimental Value  \\
\hline
$M_W$ & & $80.410 \pm 0.032$~GeV \\
$\Gamma_Z$ &  & $2.4952 \pm 0.0023$~GeV \\
$R_Z$ & & $20.767 \pm 0.025$\\
$R_b$ & & $0.21629 \pm 0.00066$ \\
$R_c$ & & $0.1721  \pm 0.0030$ \\
$A_{LR}$ & & $0.1465\pm 0.0032 $\\
$A_b$ & & $0.923\pm 0.020 $\\
$A_c$ & & $0.670 \pm 0.027$   \\
$A_{FB}^{0,l}$ & & $0.01714\pm 0.00095$ \\
$A_{FB}^{0,b}$ & & $0.0992\pm 0.0016$ \\
$A_{FB}^{0,c} $ & & $ 0.0707 \pm 0.0035 $\\
\hline
\end{tabular}
\end{center}
\caption{
\label{tab:line}
Precision data included in the global fit~\cite{lepewwg}.
}
\end{table}
The observables in the triplet Higgs model are calculated at 1-loop order 
using the results of Refs.~\cite{Blank:1997qa} and \cite{Chen:2005jx}.
Due to the presence of the extra Higgs bosons, the theoretical predictions are 
different from those of the Standard Model. 
 
The effective vector and axial vector couplings of the fermion $f$ to the $Z$ boson, 
$g_{V}^{f}$ and $g_{A}^{f}$, are determined  at one loop in the triplet model, 
\begin{eqnarray}
g_{V}^{f} & = & \biggl( \rho \frac{1-\Delta \tilde{r}}{1+\hat{\Pi}^{Z}(M_{Z}^{2})}\biggr)^{1/2}
\label{eq:gv}\\
&&
\cdot \biggl[ g_{V0}^{f} + 2 s_{\theta}c_{\theta}Q_{f}\hat{\Pi}^{\gamma Z} (M_{Z}^{2}) 
+ F_{V}^{Zf}(M_{Z}^{2})\biggr]\nonumber\\
g_{A}^{f} & = & \biggl( \rho \frac{1-\Delta \tilde{r}}{1+\hat{\Pi}^{Z}(M_{Z}^{2})}\biggr)^{1/2}
\cdot \biggl[g_{A0}^{f} + F_{A}^{Zf}(M_{Z}^{2})\biggr] \; .
\label{eq:ga}
\end{eqnarray}
They completely determine the observables of Table 1.

\begin{figure}[thb!]
\begin{center}
\hspace*{-1.5cm}
\includegraphics[scale=0.35]{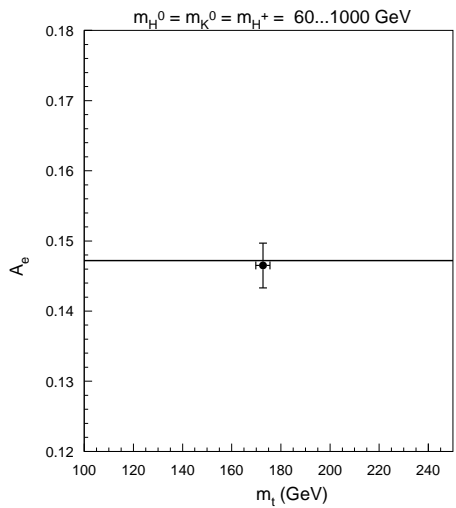}
\caption[ ]{Prediction for the left-right asymmetry, $A_{LR}=A_{e}$, 
as a function of $m_{t}$.The scalar masses, 
$m_{H^{0}}$, $m_{K^{0}}$ and $m_{H^{+}}$ are taken to be equal and 
are allowed to vary between 60 GeV to 1 TeV. The error bars on the data point 
represent the 1$\sigma$ experimentally allowed region.  
}
\label{fg:aemt}
\end{center}
\end{figure}

Using the one-loop corrected effective couplings $g_{V}^{f}$ and $g_{A}^{f}$, 
we can then calculate various $Z$-pole observables. The on-resonance 
asymmetries $A_{f}$ are determined by the effective coupling constants via the following 
relation,
\begin{equation}
A_{f}=\frac{2g_{V}^{f} g_{A}^{f}}{(g_{V}^{f})^{2}+(g_{A}^{f})^{2}} \, .
\end{equation}
Specifically, the left-right asymmetry is defined as $A_{LR}=A_{e}$, and the forward-backward asymmetry 
of the fermion as $A_{FB}^{f}=\frac{3}{4}A_{e}A_{f}$. 
The dependence of the asymmetries 
on the scalar masses and $m_{t}$ appears 
only at $\mathcal{O}((\frac{1}{16\pi^{2}})^{2})$ in the Higgs triplet model. 
 As a result, predictions for various 
asymmetries, $A_{LR}$, $A_{b}$, $A_{c}$, $A_{FB}^{0\ell}$, $A_{FB}^{0,b}$ and $A_{FB}^{0,c}$, 
are relatively {\it insensitive} to $m_{t}$ and to the 
scalar masses, $M_{H^{0}}$, $M_{K^{0}}$, and $M_{H^{+}}$. 
The prediction for the left-right asymmetry, $A_{LR}$, is shown in Fig.~\ref{fg:aemt}.

The partial width of $Z$ decay to the fermion pair $f {\overline f}$ 
 is given by~\cite{Blank:1997qa}, 
\begin{eqnarray}
\Gamma_{f} & = & \Gamma_{0} \biggl(
(g_{V}^{f})^{2} + (g_{A}^{f})^{2} \biggl( 1 - \frac{6m_{f}^{2}}{M_{Z}^{2}} \biggr)\biggr)
\\
&&
\cdot \biggl( 1 + Q_{f}^{2} \frac{3\alpha}{4\pi} \biggr)
+ \Delta \Gamma_{QCD}^{f} \; , \nonumber
\end{eqnarray}
where $\Gamma_{0} = \frac{\sqrt{2} N_{C}^{f} G_{\mu} M_{Z}^{3}}{12\pi}$, 
$N_{C}^{f} = 1 \; (3)$ for leptons (quarks), and 
$\Delta \Gamma_{QCD}^{f}$ summarizes  
the QCD corrections~\cite{Blank:1997qa}.  
Note that $g_{V}^{f}$ and $g_{A}^{f}$ are the one-loop effective 
coupling constants, determined by Eq.~\ref{eq:gv} and \ref{eq:ga}. 
The factor $( 1+ 3\alpha Q_{f}^{2}/4\pi  )$ includes the corrections to the prefactor 
in the partial decay width. 
The total $Z$ width is the sum of the fermion partial widths, 
$\Gamma_{Z} = \sum_{f} \Gamma_{f}$. 
Various ratios at the $Z$-pole are included in the fit and are defined as, 
$R_{Z} = \Gamma_{\mbox{\tiny had}}/\Gamma_{e}$, 
$R_{c} = \Gamma_{c}/\Gamma_{\mbox{\tiny had}}$, and 
$R_{b} = \Gamma_{b}/\Gamma_{\mbox{\tiny had}}$. 

\begin{figure}[thb!]
\begin{center}
\hspace*{-1.5cm}
\includegraphics[scale=0.35]{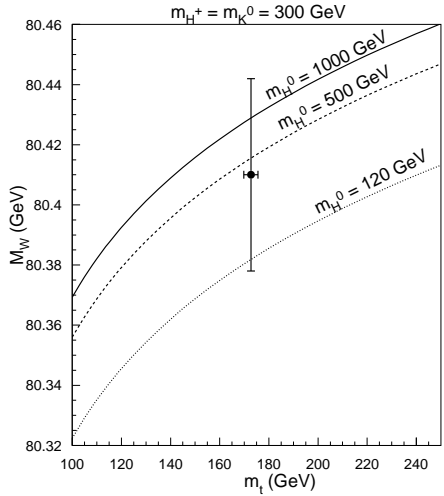}
\caption[ ]{Prediction for $M_{W}$ as a function of $m_{t}$ for $M_{H^{0}}=120$, 
$500$ and $1000$ GeV\cite{Chen:2005jx}. 
 The masses for $M_{H^{+}}$ and $M_{K^{0}}$ are 
taken to be $300$ GeV. The error bars on the data point represent the $1\sigma$ 
experimentally allowed region. 
}
\label{fg:mwmt}
\end{center}
\end{figure}

A numerical fit to just the $W$ mass 
showed that there were large cancellations between the various 
contributions in the Higgs triplet model and the lightest neutral 
Higgs boson could be heavy, as shown in Fig.~\ref{fg:mwmt}. (All
of our fits take $\gamma=0$).
This has been observed previously in generic models where 
$\rho\ne 1$ at tree level~\cite{Chen:2005jx,Blank:1997qa}. 
Furthermore, the prediction 
for $M_{W}^{2}$ exhibits a logarithmic dependence, rather than a quadratic 
one, on $m_{t}$. In other words, the $m_{t}^{2}$ dependence in the case with 
a triplet Higgs has been absorbed into the definition of 
$\sin^{2}\theta_{\mbox{\tiny eff}}$ 
(or equivalently, the definition of $\rho$.) 

\begin{figure}[thb!]
\begin{center}
\hspace*{-1.5cm}
\includegraphics[scale=0.4]{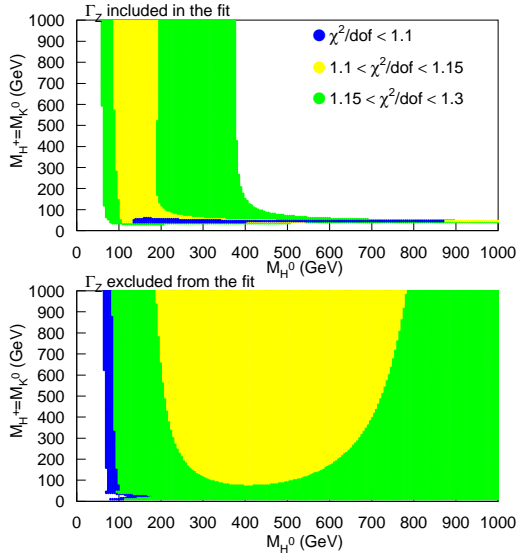}
\caption[ ]{The allowed parameter space in the $M_{H^{0}}$ and $M_{H^{+}}$ plane 
for various $\chi^{2}$ values with and without the constraint from $\Gamma_{Z}$. 
The mass $M_{K^{0}}$ is taken to be equal to $M_{H^{0}}$. The top quark mass 
is taken to be $m_{t} = 172.7$ GeV.}
\label{fg:m2h}
\end{center}
\end{figure}

\begin{figure}[thb!]
\begin{center}
\hspace*{-1.5cm}
\includegraphics[scale=0.4]{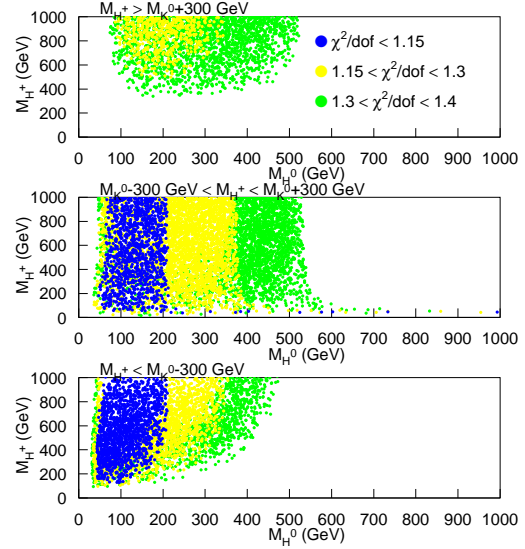}
\caption[ ]{The allowed parameter space in the $M_{H^{0}}$ and $M_{H^{+}}$ plane for 
various $\chi^{2}$ values. Here we consider mass splitting between $M_{K^{0}}$ and $M_{H^{+}}$. 
Three possible cases are considered: ({\it i}) $M_{H^{+}}-M_{K^{0}} > 300$ GeV; 
({\it ii}) $\big| M_{H^{+}}-M_{K^{0}}\big| < 300$ GeV; ({\it iii}) $M_{K^{0}} - M_{H^{+}} > 300$ GeV. 
The top quark mass is taken to be $m_{t}=172.7$ GeV. }
\label{fg:m2hier}
\end{center}
\end{figure}

It is interesting to note the pivotal role of $\Gamma_{Z}$ in the fit. 
In the global fit without including the constraint from the 
experimental value of $\Gamma_{Z}$, 
the allowed parameter space for $M_{H^{0}}$ is rather broad, 
ranging from $100$ GeV to $1$ TeV. This is consistent with the fit 
obtained to the $W$ mass alone. However, if we include the constraint 
from the $\Gamma_{Z}$ measurement, 
 the best fit to the data then occurs for a light Standard Model 
like Higgs boson with 
$100~\mbox{GeV} < M_{H^{0}} < 200~\mbox{GeV}$ and 
 degenerate $M_{H^{\pm}}=M_{K^{0}}$, 
as illustrated in Fig.~\ref{fg:m2h}. Note that the fit is not sensitive to the mass 
of the degenerate Higgs bosons. This is consistent with the reults of Ref.~\cite{Erler:2004nh}.
We have also investigated the case where a mass splitting between $M_{H^{+}}$ 
and $M_{K^{0}}$ is present, as shown in Fig.~\ref{fg:m2hier}. In this case, 
there are large contributions proportional to differences in the 
scalar masses. When the mass splitting is large, the contributions from the heavy 
scalars can be significant. 
It is found that the case in 
 which the charged Higgs is heavier than the additional 
neutral Higgs, {\it i.e.} $M_{H^{+}}\gg M_{K^{0}}$, is disfavored, while the 
cases of $M_{H^{+}}\sim M_{K^{0}}$ and $M_{H^{+}}\ll M_{K^{0}}$ are allowed. 

The reason why $\Gamma_{Z}$ plays such an important role in the global fit in 
the triplet model can be understood in the following way. 
\begin{figure}[thb!]
\begin{center}
\hspace*{-1.5cm}
\includegraphics[scale=0.3]{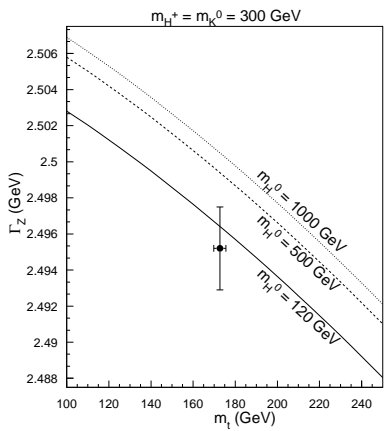}
\caption[ ]{Prediction for $\Gamma_Z$ as a function of $m_{t}$ for $M_{H^{0}}=120, \, 
500$ and $1000$ GeV. The masses for $M_{H^{+}}$ and $M_{K^{0}}$ are 
taken to be $300$ GeV. The error bars on the data point represent 
the $1\sigma$ allowed region. 
}
\label{fg:gzmt}
\end{center}
\end{figure}
The asymmetries in the triplet case 
do not receive corrections up to $\mathcal{O}((\frac{1}{16\pi^{2}})^{2})$. 
Thus the only observables in the triplet model that are sensitive to $m_{t}$ and 
$M_{H^{0}}$ are $M_{W}$ and $\Gamma_{Z}$. 
As a result, $\Gamma_{Z}$ plays an important role 
in the $\chi^{2}$ fit and can significantly constrain the allowed parameter space for $M_{H^{0}}$. 
On the other hand, all the observables considered in the global fit in the SM are 
sensitive to $m_{t}$ and $M_{H^{0}}$. The 
 constraint on $\Gamma_{Z}$ alone therefore does not have such a 
large effect. This also has the implication that the $\chi^{2}/\mbox{dof}$ value for 
the global fit in the SM is not as good as that in the triplet model. 

We comment that even though in the triplet model many observables exhibit only very 
mild logarithmic or no dependence on $m_{t}$ up to $\mathcal{O}(\frac{1}{16\pi^{2}})$, 
the $Z$-width $\Gamma_{Z}$ still depends on $m_{t}$ quadratically, 
as can be seen in Fig.~\ref{fg:gzmt}. Furthermore, in the triplet model, 
$\Gamma_{Z}$ decreases as $m_{t}$ increases, while in the SM case, 
$\Gamma_{Z}$ increases as $m_{t}$.  
Due to this strong dependence on $m_{t}$ through $\Gamma_{Z}$, it is still possible 
to place limits on $m_{t}$ using the precision data in the Higgs triplet model. 

It is interesting to note that in the littlest Higgs model with T-parity, 
a SM-like Higgs boson as heavy as 800 GeV is allowed by the global 
fit~\cite{Hubisz:2005tx} (in this case, as tree level custodial symmetry is preserved, 
a three-parameter global fit is appropriate).  
This is because the new heavy top quark which exists in this 
model gives a positive contribution to the $\rho$ parameter 
which cancels the large negative contribution from the heavy Higgs boson. 

\begin{figure}[thb!]
\begin{center}
\hspace*{-1.5cm}
\includegraphics[scale=0.35]{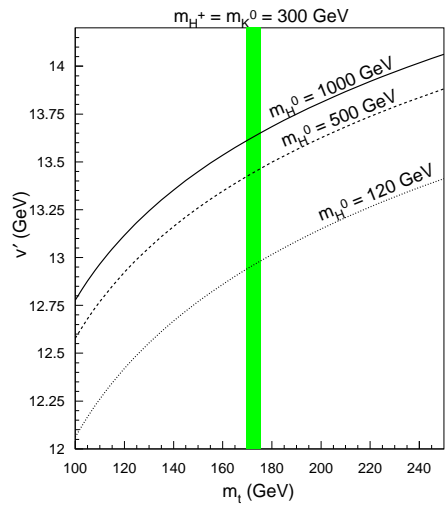}
\caption[ ]{Prediction for the triplet VEV, $v^\prime$, as a function of $m_{t}$ 
for $M_{H^{0}}=120$, $500$ and $1000$ GeV. 
 The masses $M_{H^{+}}$ and $M_{K^{0}}$ are   
taken to be $300$ GeV.  
 The band represents the $1\sigma$ 
experimentally allowed region for $m_{t}$. 
}
\label{fg:vpmt}
\end{center}
\end{figure}

\begin{figure}[thb!]
\begin{center}
\hspace*{-1.5cm}
\includegraphics[scale=0.35]{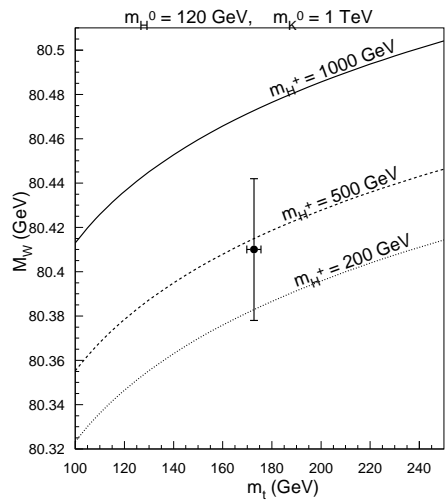}
\caption[ ]{Prediction for $M_W$ as a function of $m_{t}$ in the presence of non-degenerate scalar masses for $M_{H^{+}}=200$, 
$500$ and $1000$ GeV. 
 The mass $M_{H^{0}}$ is taken to be $120$ GeV and $M_{K^{0}}$ is  
taken to be $1$ TeV. The error bar represents the $1\sigma$ 
experimentally allowed region for $M_W$ and $m_t$. 
}
\label{fg:mwmt_nondeg}
\end{center}
\end{figure}

\begin{figure}[thb!]
\begin{center}
\hspace*{-1.5cm}
\includegraphics[scale=0.35]{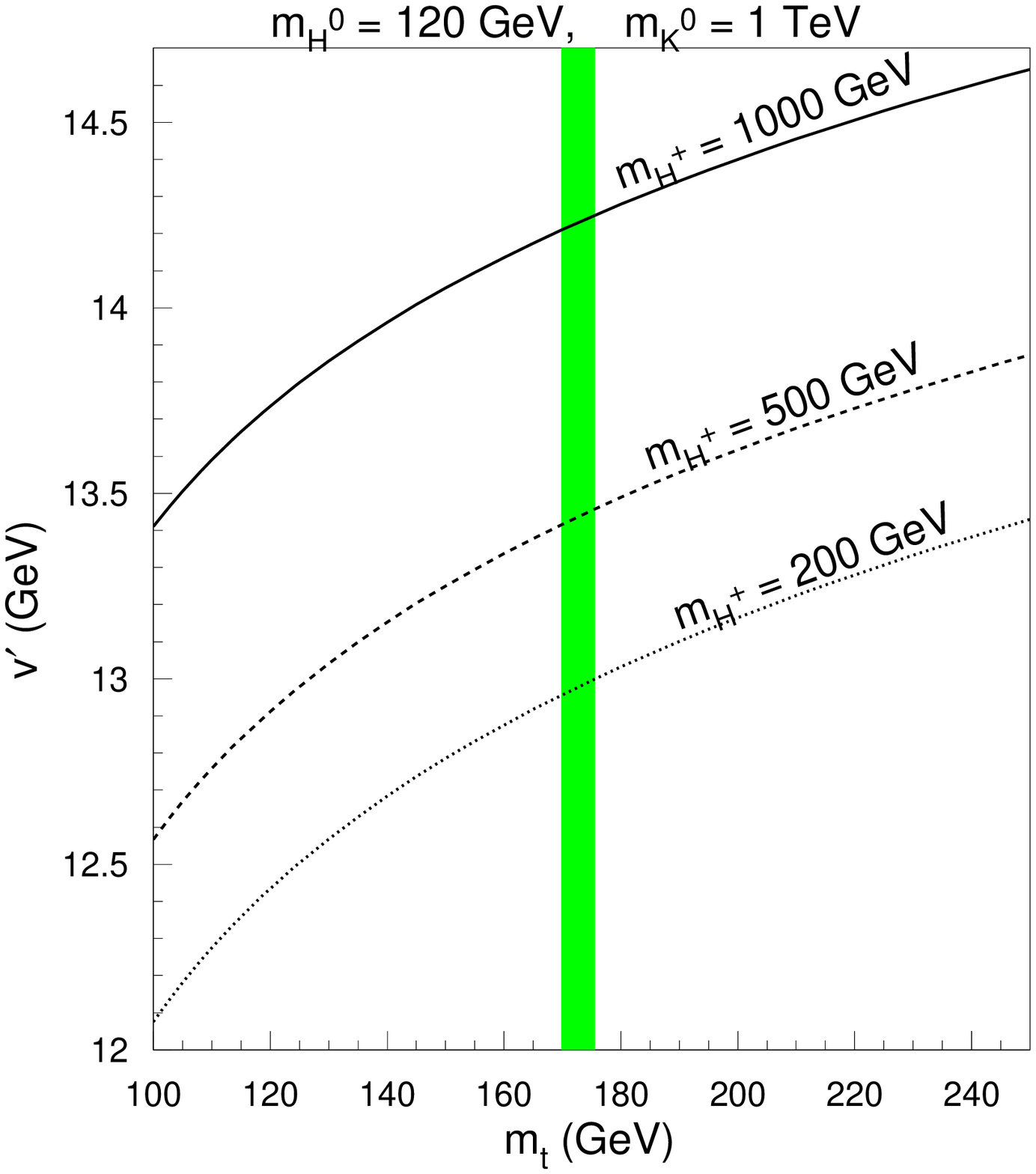}
\caption[ ]{Prediction for the triplet VEV, $v^\prime$, as a function of $m_{t}$ in the presence of non-degenerate scalar masses 
for $M_{H^{+}}=200$, $500$ and $1000$ GeV. 
 The mass for $M_{H^{0}}$ is taken to be $120$ GeV and $M_{K^{0}}$ is  
taken to be $1$ TeV. The band represents the $1\sigma$ 
experimentally allowed region for $m_{t}$. 
}
\label{fg:vpmt_nondeg}
\end{center}
\end{figure}

{\bf 4. Understanding the Triplet VEV:} 
It is interesting to interpret our results as a limit on the triplet
VEV, $v^\prime$. We calculate $v^\prime$ from the relationship,
\begin{equation}
\label{eq:delta}
\rho=1+{4v^{\prime~2}\over v^2} ={M_W^2\over M_Z^2 c_\theta^2}
= \frac{1}{\cos^{2}\delta}
\end{equation}
where $v_{SM}^2=v^2+v^{\prime~2}=(246~\mbox{GeV})^2$. We note that
$v^\prime$ depends only on $M_W$ and not on the other observables
of the global fit. In Fig.~\ref{fg:vpmt} 
we show the prediction for $v^\prime$ as a function of $m_t$ at one-loop.
By comparison with Fig.~\ref{fg:mwmt}, we see that for $M_{K^0}=
M_{H^\pm}$, the value of $v^\prime$ which correctly reproduces the experimental
value of $M_W$ ranges from $v^\prime=12.85$ GeV for $M_{H^0}=120$ GeV to
$v^\prime=13.6$ GeV for $M_{H^0}=1$ TeV. 
Figures \ref{fg:mwmt_nondeg} and  \ref{fg:vpmt_nondeg} show the predictions
for $M_W$ and $v^\prime$ for non-degenerate  $M_{K^0}$ and 
$M_{H^\pm}$, the experimental value of $M_W$ can be obtained for $v^\prime \sim
12~$ GeV. The mixing angle in the neutral Higgs sector, $\delta$, can be extracted using Eq.~\ref{eq:delta}. It is found to be 
$\delta \sim 6^{o}$. 

It is interesting to compare our results with the tree level results of 
Erler and Langacker to the parameters of the triplet model contained in 
Ref.~\cite{Eidelman:2004wy}.  In this reference they found that the best fit was obtained with $\rho \simeq \rho_{0} (1+\rho_{t}) \simeq 1.0096$, 
where $\rho_{0}$ includes the new physics contributions and $\rho_{t} = 0.00935 \left( m_{t} / 172.7~ \mbox{GeV} \right)^{2}$. 
This corresponds to $v^\prime = 12.03$ GeV for $m_{t} = 172.7$ GeV. This tree level bound on the triplet VEV, $v^\prime$, is smaller than the 
bound we found at one-loop. The difference between this result and our results can be attributed to
the important effects of the scalar loops. 


{\bf 5. Conclusion: } In models with a triplet Higgs boson, an extra 
input parameter is required for a consistent 
renormalization scheme, which significantly changes the predictions 
from those of the SM. Using the constraints on $M_{W}$ alone, 
Ref.~\cite{Chen:2005jx} found that a heavy Higgs boson 
 with a  mass as large as $\sim 1$ TeV is 
allowed in a model with a triplet Higgs. This letter contains 
our results from a global fit to 11 electroweak measurements 
in the Higgs triplet model. 
 
A large range for $M_{H^{0}}$ is allowed in the triplet Higgs model 
by all precision data except for $\Gamma_{Z}$, which 
 rules out many of the otherwise allowed values for $M_{H^{0}}$. 
As a result, a mass range of $100-200$ GeV is favored for the lightest neutral 
Higgs boson mass. This is in contrast with the SM where the Higgs mass allowed 
by the $\Gamma_{Z}$ measurement alone is heavier than the Higgs mass favored by 
the global fit \cite{lepewwg}.
A global fit for models with a triplet Higgs has been performed 
before~\cite{Erler:2004nh}, in which the allowed range for the Higgs mass is 
very close to the range we found. 
A major difference between the previous analysis presented in~\cite{Erler:2004nh} 
and our work is that the one-loop contributions from the heavy 
scalar fields were not included in the former case. 
The one-loop contributions from the heavy scalar particles could be substential 
in some parameter space, because many observables depend on scalar masses quadratically 
and the triplet Higgs does not decouple. (This non-decoupling behavior has also been observed 
in \cite{Chankowski:2006hs}. See Note Added). 

An important conclusion that should be drawn from our 
study is the importance of the one-loop contributions 
as well as the crucial role of a global fit in deriving conclusions 
about the allowed masses.
Figures \ref{fg:m2h} and 
\ref{fg:m2hier} demonstrate that the preferred value for the lightest 
neutral Higgs boson in this model
is between $100$ and $200$ GeV, as in the Standard
Model.  However, in general, the Standard Model is not the low energy
limit of the model we consider because the triplet model 
analyzed here contains
additional scalars ($K^0$ and $H^\pm$) which are allowed by the 
electroweak measurements to be as light as $200$ GeV.

{\bf Note added:} After we submitted our paper, a paper \cite{Chankowski:2006hs} 
came out in which the electroweak radiative corrections are analyzed with a different 
renormalization scheme. The non-decoupling behavior of the triplet Higgs is also observed 
in this paper.

{\bf Acknowledgments. }
We thank Paul Langacker for valuable correspondance.
Fermilab is operated by Universities 
Research Association Inc. under Contract No. DE-AC02-76CH03000 
with the U.S. Department of Energy. 
This manuscript has also been authored by Brookhaven Science Associates, 
LLC under Contract No. DE-AC02-76CH1-886 with the U.S. 
Department of Energy. The United States Government retains, 
and the publisher, by accepting the article for publication, 
acknowledges, a world-wide license to publish or reproduce 
the published form of this manuscript, or allow others to do so, 
for the United States Government purpose. 



\end{document}